\documentclass[prl, aps,twocolumn, amsmath, graphicx, latexsym, asmsymb, amsfonts]{revtex4}

\usepackage{epsfig}

\begin{document}
\title{Entanglement in doped Resonating Valence Bond states}
\author{Ravishankar Ramanathan\(^{1,2}\), Dagomir Kaszlikowski\(^{1,2}\), Marcin Wiesniak\(^{1,2}\), Vlatko Vedral\(^{1,2,3}\)}
\affiliation{
\(^1\) Centre for Quantum Technologies, National University of Singapore, 117543 Singapore\\
\(^2\) Department of Physics, National University of Singapore, 117542 Singapore\\
\(^3\)The School of Physics and Astronomy, University of Leeds, Leeds, LS2 9JT, United Kingdom}

\date{\today}

\begin{abstract}
We investigate the entanglement properties of resonating valence bond states on a two dimensional lattice in the presence 
of dopants that remove electrons from the lattice creating holes. The movement of the holes generated by the Hubbard Hamiltonian in the 
regime of strong Coloumb repulsion in this setting could be responsible for the phenomenon of high temperature superconductivity as hypothesised
by Anderson in Science {\bf 235}, 1196, (1987). We argue that there is a particular density of dopants (holes) where the entanglement 
contained in the lattice attains its maximal value for the nearest-neighbour RVB liquid state.
\end{abstract}

\maketitle

\section{Introduction.}
In quantum many-body physics, resonating-valence-bond (RVB) states have received a lot of attention due to their importance in 
the description of different 
phenomena. 
They are used to describe the 
resonance of covalent bonds in organic molecules, behavior of Mott insulators without long-range  antiferromagnetic order \cite{anderson}, 
superconductivity in organic solids \cite{organic-solids}, and the recently discovered 
insulator-superconductor transition in boron-doped diamond \cite{boron}. There are many other applications of RVB states (see e.g. \cite{review}). 
Moreover, RVB states have been suggested as a basis for fault-tolerant topological quantum computation \cite{kitaev}. 

It was postulated by Anderson in Ref. \cite{anderson} that the short range nearest-neighbour RVB state (also called RVB liquid) might be responsible for the phenomenon of high temperature 
superconductivity. The cuprate superconductors are recognized as doped Mott insulators as in the case of Strontium doped Lanthanum cuprate LSCO. The copper oxide planes are described by the one-band Hubbard model with a strong, on-site repulsion U. 
The RVB state on a square lattice for the Hubbard Hamiltonian with strong Coloumb repulsion was proposed by Anderson
to be the Mott insulator phase of the system, i.e., the pure Lanthanum cuprate is in an RVB state. 
By introducing dopants one removes some electrons from the lattice creating "holes", i.e., unoccupied
sites on the lattice. As pointed out by Anderson, it would take a finite concentration of dopants to metallize the material and make it a superconductor as initially the dopants will be screened by the bound quasiparticles. The holes start to hop from one site to another and their motion is uninhibited leading to their delocalization over the whole lattice, which 
can be interpreted as a persistent current. The movement of the holes is resistant to thermal noise and depends on the density $\mu$ of dopants (holes)
reaching its maximum somewhere between $10\%$ to $15\%$ (the so-called 1/8 anomaly) of holes on the lattice. It is notable that there does not seem to be a clear consensus on the origin of this anomaly.

Indeed, the ground state of the 2D Hubbard model with doping is still unknown.
Numerical simulations indicate that the RVB scenario is the right one for coupled plaquettes and ladders \cite{Fye}, and recently experiments have been proposed to test the RVB scenario in fermionic atoms in 2D optical lattices \cite{zoller}. These experiments propose methods to increase the inter-ladder coupling to check if the RVB state on ladders is adiabatically connected to the Hubbard model ground state on the square lattice, which would provide an experimental test of the RVB theory. In this paper, we investigate the entanglement properties of the RVB states in small-sized laddders and plaquettes and speculate on the behavior in the thermodynamic limit.

The entanglement properties of RVB states without dopants on many dimensional lattices has been investigated in \cite{anushya}. The main
conclusion of the Ref. \cite{anushya} is that such states can only have (if any) a very small amount of bipartite entanglement between any two sites on the lattice
but they are always genuinely multi-partite entangled. 

In view of this, it is interesting to see how multipartite entanglement of RVB states depends on the density of holes. In particular, we are interested if 
changes in the amount of entanglement correspond to the experimental observation of the maximal Tc superconductivity in the hole density window 
$0.1<\mu<0.15$. Intuitively, one would expect entanglement to be larger in this window as well. This is because entanglement and supercurrent are both related to the existence of correlations and therefore their peaks should be related i.e. a larger entanglement would be more robust to increase in temperature.

To address these questions we first define the RVB states on a $2 L$ site lattice with $2 n$ holes and an appropriate entanglement measure. Subsequently, we investigate analytically lattices up to 24 sites with open boundary conditions. Based on the obtained results we 
conjecture that in the thermodynamic limit of $L\rightarrow\infty$ the amount of entanglement reaches its maximum for some critical density of holes $\mu_{cr}$, where $0<\mu_{cr}<1.$

\section{Formulation of the problem}
Let us consider a two dimensional lattice with open boundary conditions consisting of $2L$ sites, which is a union of two sub-lattices $A$ and $B$ in such a way
that any site belonging to the sub-lattice $A(B)$ has all its nearest neighbours belonging to the sub-lattice $B(A)$. We define a {\it dimer} between
sites $a\in A$ and $b\in B$ as a singlet state $|\delta_{ab}\rangle = \frac{1}{\sqrt{2}}(|0\rangle_a|1\rangle_b-|1\rangle_a|0\rangle_b)$. A {\it dimer covering}
is defined as a tensor product of dimers $|\Delta_{[(a_1b_1),(a_2b_2),\dots, (a_{L} b_{L})]}\rangle = \otimes_{k=1}^{L}|\delta_{a_kb_k}\rangle$.
Here the set of pairs $[(a_1b_1),(a_2b_2),\dots, (a_{L} b_{L})]$ represents a particular way of joining neighbouring sites of the two sub-lattices with singlets. 
The number of such sets for the case of square lattices with open boundary conditions is, from \cite{Fisher}, \cite{Kastelyn2}, given by $\prod^{\sqrt{\frac{L}{2}}}_{j=1}\prod^{\sqrt{\frac{L}{2}}}_{k=1}(4 \cos^{2}(\frac{\pi j}{2n+1}) + 4 \cos^{2}(\frac{\pi k}{2n+1}))$.
For periodic boundary conditions in the square lattice this number is known to be $(\exp{(\frac{2 G}{\pi})})^L\approx(1.791)^L$, where $G = \sum^{\infty}_{n=0}(-1)^n/(2n+1)^2$ is Catalan's constant.

The RVB liquid is defined as 
\begin{eqnarray}
|\Delta\rangle = \frac{1}{\sqrt{R}}\sum_{(a_1b_1)\dots (a_Lb_L)}|\Delta_{[(a_1b_1),(a_2b_2),\dots, (a_{L} b_{L})]}\rangle,
\end{eqnarray} 
where $R$ is the normalization constant and the summation extends over all possible dimer coverings. $R$ can be in principle calculated using the
techniques of the so-called 'random loop soup' \cite{tasaki}. In practice, the problem of counting is analytically intractable and even with the help
of numerics one cannot compute $R$ for large $L$. 

 
Let us now denote an equal superposition of dimer coverings of the lattice where two arbitrary sites $a_i$ and $b_j$
are unoccupied by $|\Delta_{(a_ib_j)}\rangle$. Please note that we do not allow two sites belonging to the same sub-lattice to
be unoccupied as this will preclude the possibility of covering the rest of the lattice with nearest-neighbour dimers. This guarantees that the occupied 
part of the lattice is in the RVB liquid state with the number of dimer coverings depending on the positions of the holes. 

In general, if there are $2n$ holes one can define in the same manner the state $|\Delta_{[(a_1b_1),(a_2b_2),\dots, (a_{n} b_{n})]}\rangle$, i.e.,
the superposition of all possible coverings of the part of the lattice excluding the empty sites $(a_1b_1),(a_2b_2),\dots, (a_{n} b_{n})$.
Note that if $n$ is too large it may not be possible to cover the part of the lattice without holes by dimers. A simple example is
when $2n=L$. In this situation if every second site is empty (chess board configuration), the remaining part of the lattice cannot be covered by dimers. 
We will address this issue further on.

The final steady state of the system, i.e., the state after the system of the lattice and dopants has reached an equilibrium can be written then as 
\begin{eqnarray}
&&|\Delta_{2n}\rangle = \sum_{(a_{i_1}b_{j_1}),\dots (a_{i_n}b_{j_n})}\nonumber\\
&&\sqrt{p(a_{i_1},b_{j_1},\dots a_{i_n},b_{j_n})}|\Delta_{[(a_{i_1}b_{j_1}),\dots (a_{i_n}b_{j_n})]}\rangle,
\end{eqnarray}
where the probability distribution of the holes $p(a_{i_1},b_{j_1},\dots a_{i_n},b_{j_n})$ depends on their detailed dynamics.  

We would like 
to be very clear about the meaning of $|\Delta_{2n}\rangle$, namely that, it may not be an accurate description of the state of the lattice of a real doped superconductor.
First of all, the RVB theory is one of many theories of the discussed phenomenon \cite{Anderson2}. Secondly, even within the RVB theory itself it is not clear how the state with $2n$ holes looks like \cite{Anderson3}. For instance, one cannot exclude the possibility that long-range dimers will appear in $|\Delta_{2n}\rangle$. However, it seems reasonable 
to assume that for a small amount of holes  $|\Delta_{2n}\rangle$ is an acceptable choice. 

It is not clear how to quantify multiparty entanglement in the $|\Delta_{2n}\rangle$ state. The problem arises from the fact that a commonly accepted definition of multiparty entanglement
is that any bipartition of the considered many particle quantum state must be entangled. However, in our case any bipartition of the lattice (equivalently the bipartition of the 
state $|\Delta_{2n}\rangle$) leads to a state with a variable number of particles on each site of the bipartition. According to super-selection rules one cannot observe a superposition
of states with different number of particles \cite{Wick}, which considerably complicates the task of quantifying entanglement in RVB states with holes. 

As a measure of the amount of non-classical correlations in the lattice we take the geometric measure of entanglement \cite{geoment}, which is generalized to the multi-partite case in a straight-forward manner \cite{geomentmulti}. For pure states, the measure is given by the $\frac{1}{2}$-based logarithm of the squared modulo of the overlap between the state, and the separable state closest to it
\begin{equation}
E(|\psi\rangle)=-\max_{|\psi_{sep}\rangle}|\log_2|\langle \psi|\psi_{sep}\rangle|^2,
\end{equation}
where $|\psi_{sep}\rangle$ is a separable state.
In the case of our lattice, however, we deal with the subtle matter of super-selection rules, because the closest product state to the RVB state could involve forbidden local superpositions of a hole and an electron. For this reason, we use the average geometric measure. The averaging is done over all possible locations of the holes. 

More precisely, we define {\it an average geometric measure of entanglement } on the state $|\Delta_{2n}\rangle$ as
\begin{eqnarray}
&&\bar{E}(2n)=\sum_{(a_{i_1}b_{j_1}),\dots (a_{i_n}b_{j_n})}\nonumber\\
&&p(a_{i_1},b_{j_1},\dots a_{i_n},b_{j_n})E(|\Delta_{[(a_1b_1),\dots, (a_{n} b_{n})]}\rangle),
\end{eqnarray}
where 
\begin{eqnarray}
&&E(|\Delta_{[(a_1b_1),(a_2b_2),\dots, (a_{n} b_{n})]}\rangle) =\nonumber\\
&& -2\log_2{\max_{|\psi_{sep}\rangle}{|\langle\psi_{sep}| \Delta_{[(a_1b_1),(a_2b_2),\dots, (a_{n} b_{n})]}\rangle  |}}. 
\end{eqnarray}
The maximum is taken over all fully separable states on the part of the lattice without holes and the physical meaning of $\bar{E}(2n)$ is clear; it is the average amount of entanglement one gets after locating the position of the holes on the lattice.

\subsection{RVB state without holes: $\bar{E}(0)$.}

First we consider the RVB state without holes. We already know from 
the Ref. \cite{anushya} that it contains negligible two-site entanglement but
it is genuinely multi-party entangled. Here we calculate $\bar{E}(0)$ that will serve 
us later as a basis for comparison with $\bar{E}(2n)$. Additionally, the same method of calculation will be used for $n>0$.

We have 
\begin{equation}
\bar{E}(0) = -2\max_{|\psi_{sep}\rangle}\log_2{|\langle \psi_{sep}|\Delta\rangle|} = -2\log_2{\frac{C}{R}},
\end{equation}
where $C$ is the number of dimer coverings for $|\Delta\rangle$, i.e., $C=\prod^{\sqrt{\frac{L}{2}}}_{j=1}\prod^{\sqrt{\frac{L}{2}}}_{k=1}(4 \cos^{2}(\frac{\pi j}{2n+1}) + 4 \cos^{2}(\frac{\pi k}{2n+1}))$ for open boundary conditions and $C=(\exp{(\frac{2 G}{\pi})})^L$ for periodic boundary conditions in the square lattice. 
This result can be argued as follows. 
The state $|\Delta\rangle$ always has an "anti-ferromagnetic" term of the form $|0101\dots 01\rangle$ or $|1010\dots10\rangle$
with its coefficient equal to $\pm \frac{C}{R}$. Naturally every other term has smaller coefficient, because in the superposition of all coverings only the antiferromagnetic terms add up. Thus, there is a fully separable state
$|\psi_{sep}^{(0)}\rangle = |0101\dots 01\rangle$ for which the modulus of the scalar product with $|\Delta\rangle$ equals $\frac{C}{R}$.
However, the only fully separable state with equal number of zeros and ones is of the form $|x_1x_2\dots x_n\rangle$ with $x_1+x_2+\dots
+x_n = L$, which means that $|\psi_{sep}^{(0)}\rangle = |\psi_{sep}^{(max)}\rangle$.

\subsection{RVB with $2n$ holes: $\bar{E}(2n)$.}

To compute entanglement in this case it suffices to find the maximal overlap between $|\Delta_{[(a_1b_1),(a_2b_2),\dots,(a_n b_n)]}\rangle$ and a fully separable state
for every possible set of pairs $(a_1b_1),(a_2b_2),\dots,(a_nb_n)$. From the previous considerations we know that the maximum is reached for an anti-ferromagnetic separable state
and it reads $\frac{    C_{   [(a_1b_1),(a_2b_2),\dots, (a_n b_n)]}     }{R_{[(a_1b_1),(a_2b_2),\dots, (a_{n} b_{n})]}}$, where $C_{[(a_1b_1),(a_2b_2),\dots, (a_{n} b_{n})]}$ is the number of the coverings of the initial lattice with the sites 
$(a_1,b_1),(a_2,b_2),\dots,(a_n,b_n)$ removed and $R_{[(a_1b_1),(a_2b_2),\dots, (a_{n} b_{n})]}$ is the normalization of the state $|\Delta_{[(a_1b_1),(a_2b_2),\dots, (a_{n} b_{n})]}\rangle$. Therefore, we have 
\begin{eqnarray}
&&\bar{E}(2n) = \sum_{(a_1b_1),(a_2b_2),\dots, (a_{n} b_{n})} p(a_{1},b_{1},\dots a_{n},b_{n})\nonumber\\
&& \log_2{\left(\frac{   C_{   [(a_1b_1),(a_2b_2),\dots, (a_n b_n)]}     }{R_{[(a_1b_1),(a_2b_2),\dots, (a_{n} b_{n})]}}\right)^{-2}}.
\end{eqnarray}

The main difficulty in the above formula is that it is an $NP$-complete problem to calculate $C_{   [(a_1b_1),(a_2b_2),\dots, (a_n b_n)]} $ and $R_{[(a_1b_1),(a_2b_2),\dots, (a_{n} b_{n})]}$. This stems from the known result in theoretical computer science \cite{Jerrum} that counting dimer coverings of a planar lattice is a polynomial-time computable problem, whereas counting monomer-dimer arrangements on the two-dimensional lattice is an $NP$-complete problem. In our language, the holes correspond to the monomers and the singlet pairs correspond to the dimers. Thus, finding $C_{   [(a_1b_1),(a_2b_2),\dots, (a_n b_n)]} $ is an $NP$-complete problem. Moreover, calculating $R_{[(a_1b_1),(a_2b_2),\dots, (a_{n} b_{n})]}$ is at best polynomial-time reducible to finding $C_{   [(a_1b_1),(a_2b_2),\dots, (a_n b_n)]} $.

\section{Results}

In this section we present the main results of the paper. 


We have analytically computed entanglement $\bar{E}(2n)$ using the method described above for ladders and plaquettes up to size 6x4 (ladder lattices of size $3\times2, 4\times2,\dots,10\times2$ and rectangular lattices $4\times3, 4\times4, 5\times4$ and $6\times4$). The results, shown in the Fig. [1] for ladders and Fig[2] for rectangular lattices, clearly show that $\bar{E}(2n)$ reaches the maximum at a certain hole density as the size of the lattice increases. Moreover, the maximum occurs at a low concentration of holes. 
\begin{figure}
\begin{center}
\includegraphics[width=0.4\textwidth,height=0.35\textwidth]{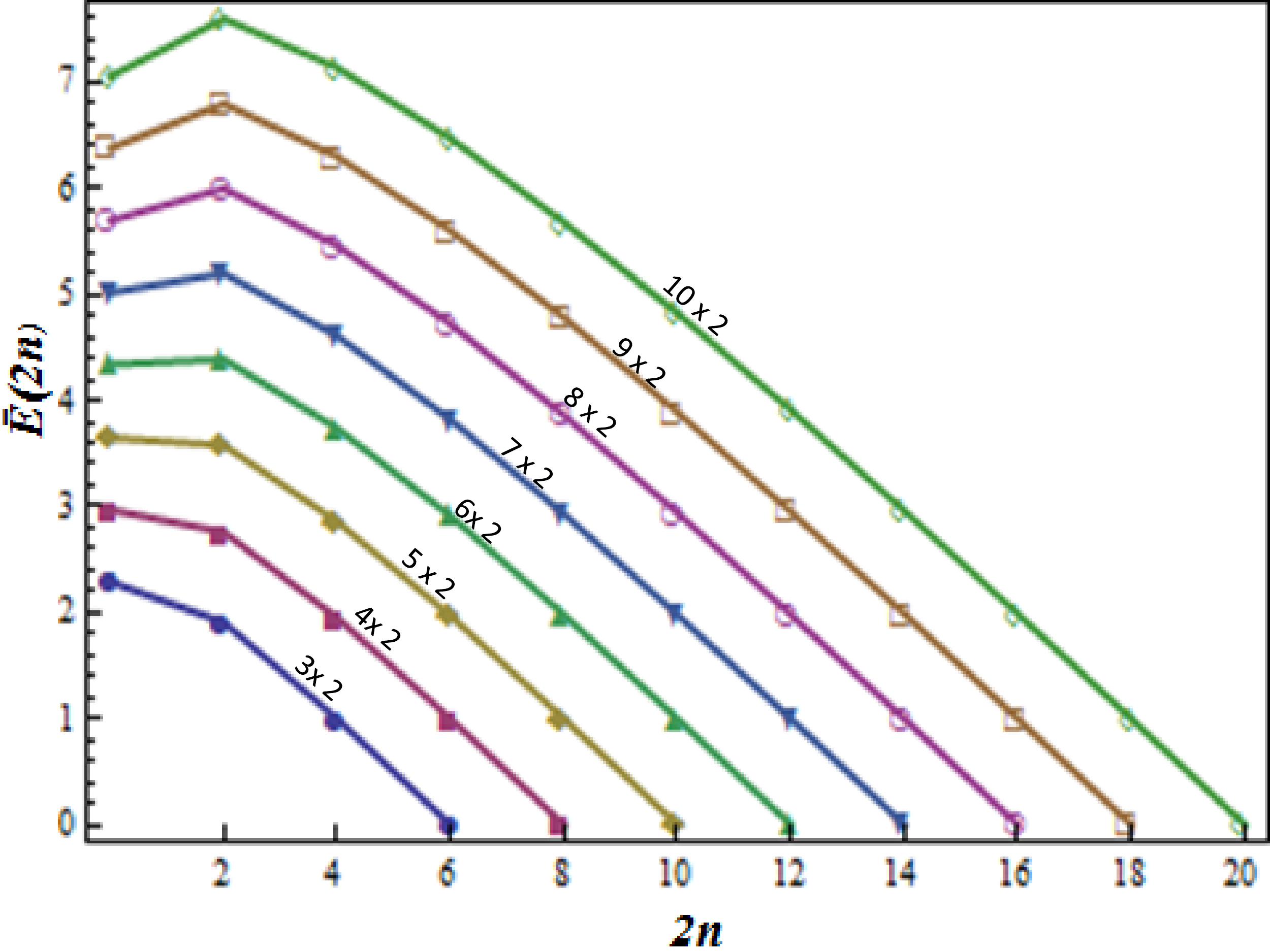}
\end{center}
\caption{$\bar{E}(2n)$ for Ladder lattices 3x2, 4x2, \dots, 10x2. Note the appearance of the peak at 2 holes for lattices larger than 6x2} \label{fig:main}
\end{figure}

\begin{figure}
\begin{center}
\includegraphics[width=0.4\textwidth,height=0.35\textwidth]{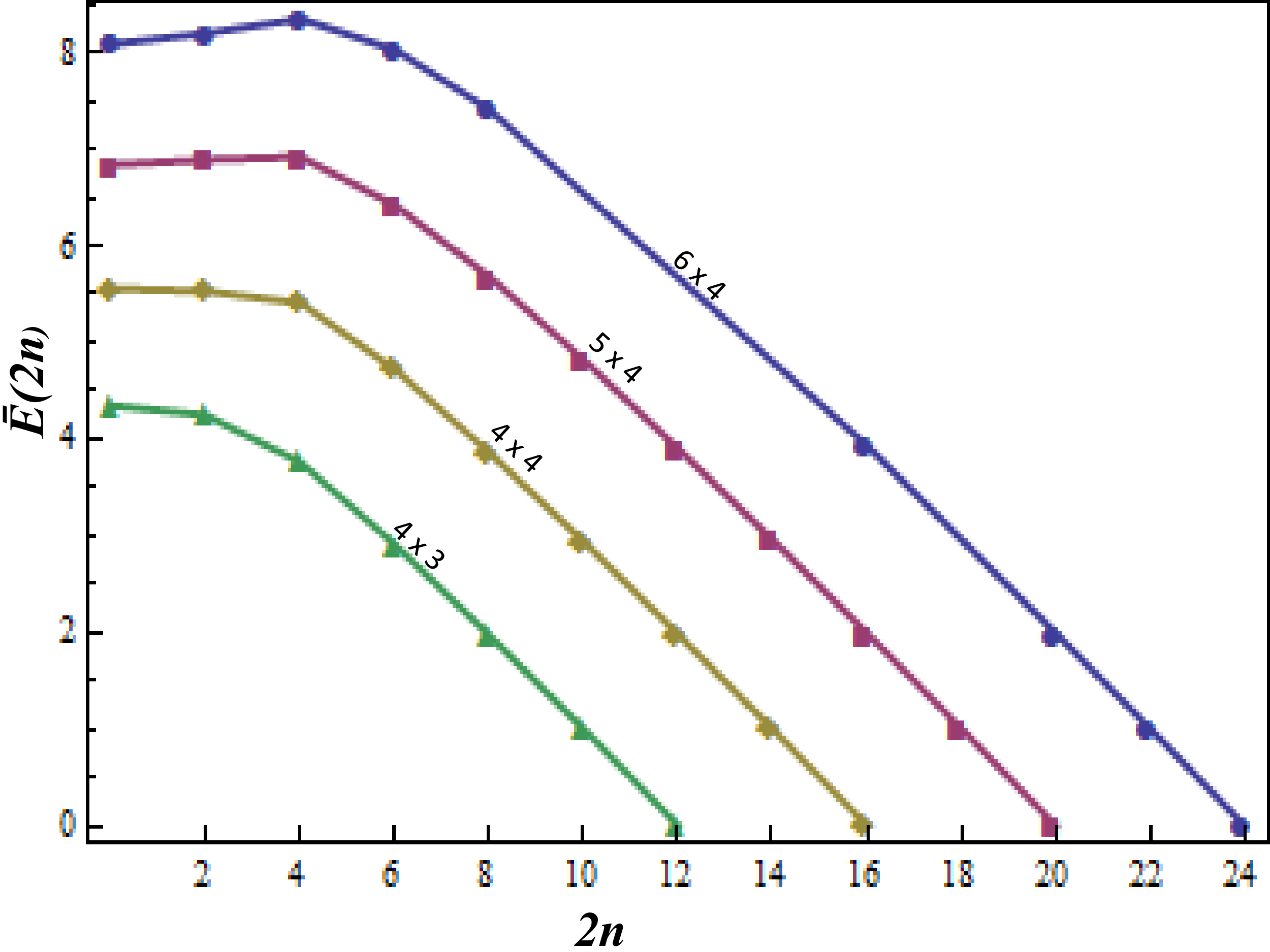}
\end{center}
\caption{$\bar{E}(2n)$ for Rectangular lattices 6x4, 5x4, 4x4, 4x3. Note the appearance of the peak at 4 holes in 5x4 and 6x4. The position of the peak has shifted to the right (higher density) compared to that in Fig [1].} \label{Fig[2]}
\end{figure}

We now elucidate certain significant points in the calculations leading to the graphs in Figs [1] and [2]. As an illustrative example, let us consider the lattice of size 6x4 with four holes, at which point the peak occurs in this structure. It is clear that the 4 holes can be in one of $C^{12}_2 \times C^{12}_2$ positions, where $C^m_n$ denotes the binomial coefficient. Hence, one has to average over the entanglement found in each of these cases to find $\bar{E}(4)$ for this lattice. However, in this calculation we omit the pathological positions of the holes in which a single site is surrounded on all sides by holes as in Fig.[3], in which case the rest of the lattice is unable to form a short-range RVB structure. Long-range dimers between sites belonging to the same sublattice would be needed to fill the lattice in such a situation and we omit the corrections accruing due to these. In any case, neglecting these situations cannot substantially alter the behaviour of entanglement because they occur with the probability of $4 (C^{(ab/2)-1}_2-1)/(C^{(ab/2)}_2)^2$ for four holes in an $a \times b$ lattice. 
Note that the probability of occurrence of these situations is zero for two holes so that the positive gradient at the beginning of the curve is maintained and the existence of the peak is assured. Thus, we conclude that the peak in the graph is maintained even when these situations involving long-range dimers are taken into consideration.

\begin{figure}
\begin{center}
\includegraphics[width=0.4\textwidth,height=0.2\textwidth]{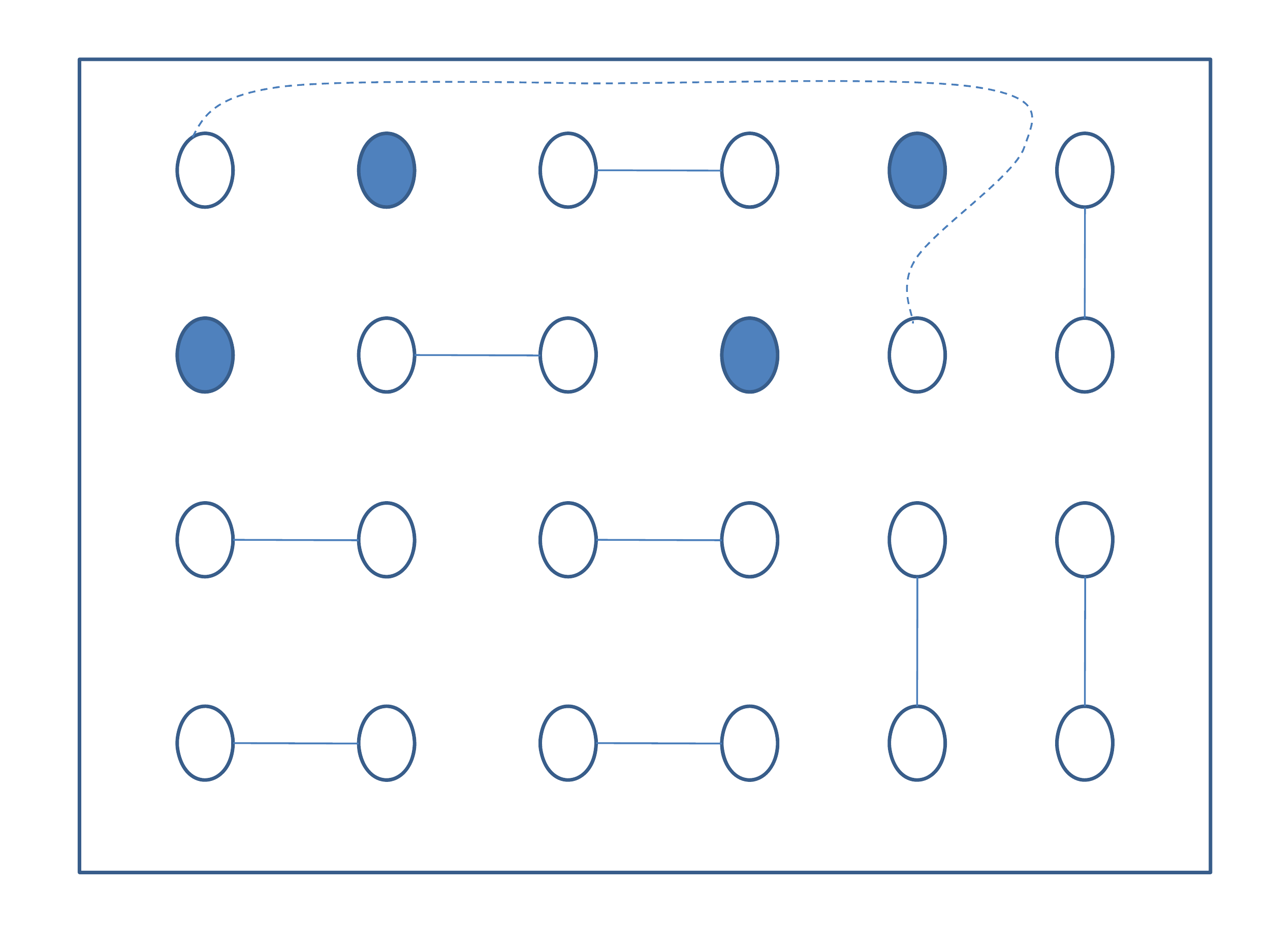}
\end{center}
\caption{Typical pathological case which is omitted from short-range RVB calculations. These cases do not affect the behavior of entanglement as explained in the text.} \label{Fig[3]}
\end{figure}

It is seen from Fig[2] that the initial gradient of the curve increases with the size of the lattice. If the trend continues for larger lattices, one might expect the peak to shift to the right and converge to a particular concentration in the thermodynamic limit. 





\section{Conclusions}

Our data strongly indicates that the average multiparty entanglement quantified by $\bar{E}(2n)$ reaches a maximum for some critical density of holes in the thermodynamical limit. Although we are unable to predict the exact value of this critical density due to the hard computational nature of the problem, we conjecture that it is located in the region of low density of holes. 

A possible way to get some additional information about the location of the maximum  in the thermodynamic limit would be to translate the problem in graph theoretic language in the following way \cite{Kastelyn}. It is apparent that each dimer covering of the RVB structure can be represented as a balanced bipartite graph with the sublattices A and B forming the two vertex sets, there being L edges in a 2L site lattice. In such a graph, there is no path between any two sites belonging to the same sublattice, and the matching number of the graph is equal to L. These conditions can be extended to the case of the lattice with holes as well. 

\begin{figure}
\begin{center}
\includegraphics[width=0.4\textwidth,height=0.35\textwidth]{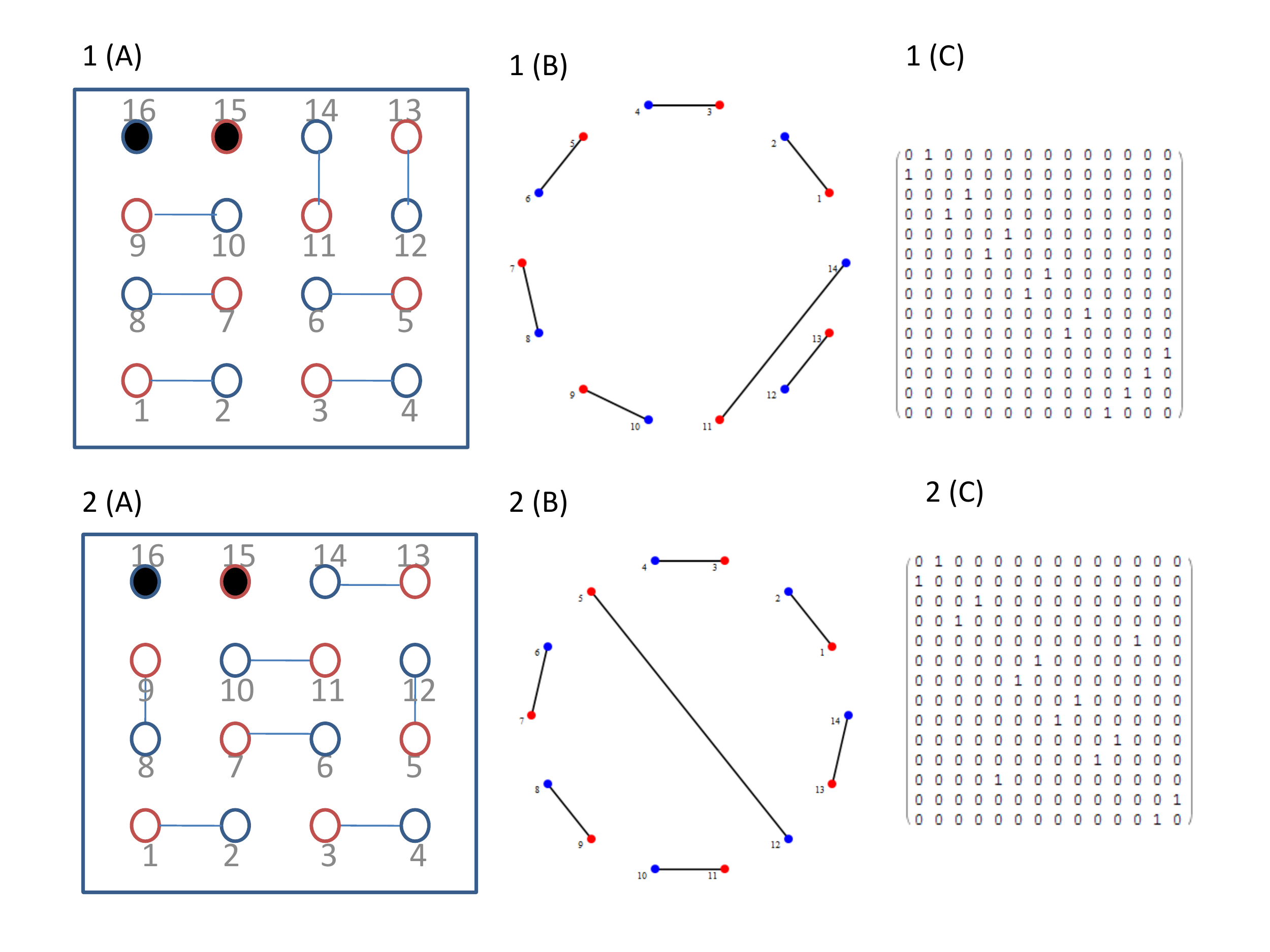}
\end{center}
\caption{Graph theoretic formulation: 1(A),2(A)- Sample coverings 1 and 2 of $4 \times 4$ lattice with two holes; 1(B),2(B)-Corresponding bipartite graphs; 1(C),2(C) Corresponding adjacency matrices to the two coverings} \label{Fig[4]}
\end{figure}

\begin{figure}
\begin{center}
\includegraphics[width=0.4\textwidth,height=0.35\textwidth]{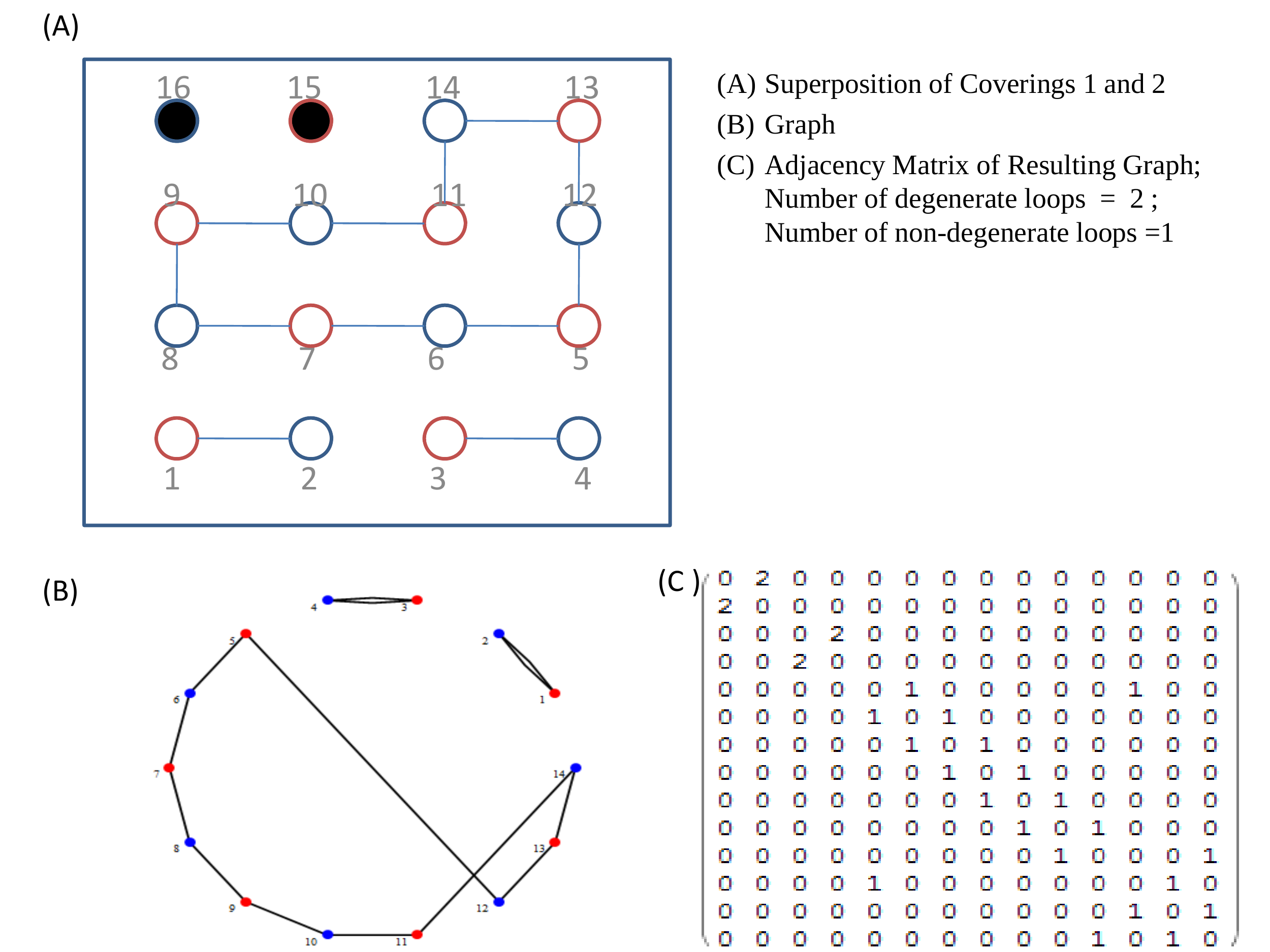}
\end{center}
\caption{Graph theoretic formulation: Coverings 1 and 2 superimposed on each other} \label{Fig[5]}
\end{figure}

The problem of finding C then translates to the equivalent problem of finding the number of graphs with adjacency matrices having the following property. Each adjacency matrix is a sparse matrix of size 2L by 2L, the mth row has a 1 at one of upto four possible positions, these corresponding to the sites adjacent to site m in the lattice; and each row and each column has only one non-zero entry. The number of such matrices then equals the number of coverings. Algorithms for approximating C have been devised as enumerating the number of perfect matchings in such graphs in \cite{Kenyon}.

The problem of finding R can also be broken down into the equivalent problem of finding the number of degenerate and non-degenerate loops in every distinct superposition of two coverings, as in \cite{Kohmoto}. To do this, we add all the adjacency matrices for that lattice, two at a time, keeping only the distinct results. The number of degenerate loops (dl) in each such superposition is then simply half the number of twos in the resulting matrix. The number of nondegenerate loops (ndl) is equal to the total number of cycles in the graph for which the resulting matrix forms the adjacency matrix, minus the number of degenerate loops. R is then found from the neat formula,\cite{Kohmoto}
\begin{equation}
 R = \sum_{superpositions}{2^{dl}\times4^{ndl}}
\end{equation}
However, there do not seem to be good approximation schemes to bind R using the above method. 

Figs[4] and [5] illustrate the method in graphic detail showing two coverings for a $4 \times 4$ lattice with two holes. In Fig[4] 1(A), the red circles indicate sublattice A sites and blue circles indicate those belonging to sublattice B. Shaded circles indicate holes in the lattice and singlets are represented as lines between sites. The bipartite graph corresponding to the state is shown to its right in 1 (B). The graph has bipartition (A,B) with $A = (1,3,5,7,9,11,13)$ and $B=(2,4,6,8,10,12,14)$. Since, $\left|A\right| = \left|B\right| = L$, the graph is balanced. Edges of the graph connect vertex set A to B such that an edge connects a vertex to only one of its nearest neighbouring vertices on the lattice. Since there are L such edges, the size of the maximum matching of the graph is equal to L. The adjacency matrix for this graph is shown alongside in 1(C). To find R, we would need to superimpose the two coverings on each other as shown in Fig [5]. The resulting graph and its adjacency matrix (the sum of the two adjacency matrices in Fig [4]) are shown alongside. The number of degenerate and non-degenerate loops can then be calculated from the number of cycles in the graph.

It is hoped that with these methods and by experimental observations as suggested in \cite{zoller} the entanglement vs hole density curve can be constructed in the thermodynamic limit. This might throw more light on the question of whether multipartite entanglement, defined in this average geometric sense could indicate the occurrence of the quantum phase transition.

\begin{acknowledgments}
We would like to thank Aditi Sen, Ujjwal Sen, Jacob Dunningham, Miklos Santa and G. Baskaran for useful discussions and comments. This work was supported by the National Research Foundation and Ministry of Education, Singapore.    

\end{acknowledgments}


\begin{thebibliography}{99}
\bibitem{anderson} P.W. Anderson, Science {\bf 235}, 1196, (1987).

\bibitem{organic-solids} T. Ishiguro \emph{et al.},
\emph{Organic Superconductors}
 (Springer, New York, 1998); J.Y. Gan \emph{et al.},
Phys. Rev. Lett. {\bf 94}, 067005 (2005); J. Liu \emph{et al.},
Phys. Rev. Lett. {\bf 94},
\emph{ibid.} 127003 (2005); B.J. Powell and R.H. McKenzie, 
\emph{ibid.}
Phys. Rev. Lett. \textbf{94},
047004 (2005).
\bibitem{boron} 
E.A. Ekimov et al., 
Nature {\bf 428}, 542 (2004).

\bibitem{review} P.A. Lee \emph{et al.},
Rev. Mod. Phys. \textbf{78}, 17 (2006); G. Baskaran, cond-mat/0611553, and 
references therein.
\bibitem{kitaev} A.Y. Kitaev, Ann. Phys. {\bf 303}, 2 (2003).
\bibitem{Fye} R.M. Fye \emph{et al.}, Phys. Rev. B \textbf{46}, 8667 (1992); Tsunetsugu \emph{et al.}, Phys. Rev. B \textbf{49}, 16078 (1994); Troyer \emph{et al.} Phys. Rev. B \textbf{53}, 251 (1996). 
\bibitem{zoller} S. Trebst \emph{et al.},
Phys. Rev. Lett. \textbf{96}, 250402 (2006); 
A.M. Black-Schaffer and S. Doniach, cond-mat/0612158, and 
references therein.
\bibitem{anushya} A. Chandran \emph{et al.},Phys. Rev. Lett. {\bf 99}, 170502 (2007); D. Kaszlikowski \emph{et al.} Phys. Rev. A {\bf 76} 054302 (2007).


\bibitem{Fisher} M.E. Fisher, Phys. Rev. {\bf 124}, 6 (1961).
\bibitem{Kastelyn2} P.W. Kastelyn, Physica \textbf{27}, 1209 (1961)
\bibitem{tasaki} H. Tasaki, Phys. Rev. B \textbf{40}, 9183 (1989). 
\bibitem{Anderson2} P.W.Anderson Phys. Scr. \textbf{T102} 10-12 (2002).
\bibitem{Anderson3} P.W.Anderson \emph{et al.}, J Phys. Condens. Matter \textbf{16} R755-R769 (2004).
\bibitem{Wick}G. Wick \emph{et al.} Phys. Rev. \textbf{88(1)} 101-105 (1952);G.Wick \emph{et al.} Phys. Rev. D \textbf{1 (12)} 3267-3269 (1970).
\bibitem{geoment} A. Shimony, {\em Ann. NY. Acad. Sci.} {\bf 755}, 675 (1995); H. Barnum and N. Linden, {J. Phys. A: Math. Gen.} {\bf 34}, 6787 (2001).
\bibitem{geomentmulti} T.-C. Wei and P.M. Goldbart, {\em Phys. Rev. A} {\bf 68}, 042307 (2003).

\bibitem{Jerrum} M. Jerrum, J. Stat. Phys. {\bf 48},1-2 (1987).
\bibitem{Kastelyn} P.W. Kastelyn, in \textit{Graph Theory and Theoretical Physics}, (Academic, London, 1967).
\bibitem{Kenyon} C. Kenyon \emph{et al.} J. Stat. Phys. \textbf{83}, 637 (1996).
\bibitem{Kohmoto} M. Kohmoto and Y. Shapir, Phys. Rev. B {\bf 37}, 16 (1988).




 \bibitem{Mermin} N.W. Ashcroft and N.D. Mermin, \emph{Solid State Physics} (Holt, Rinehart, and Winston, Philadelphia, 1979); 
\bibitem{Mott}  N. F Mott, \emph{Metal-insulator transitions} (Taylor and Francis, London, 1974).

\bibitem{schluter} M. Schluter, in \emph{Superconductivity and Applications}, ed. H.S. Kwok et al. Plenum, New York, 1990.




 











 




\end{thebibliography}
\end{document}